\documentclass[10pt,aps,prl,twocolumn,amsmath,amssymb,footinbib,showpacs,longbibliography,superscriptaddress]{revtex4-2}

\usepackage{times}

\usepackage[english]{babel}
\usepackage{latexsym}
\usepackage{graphics}
\usepackage{subfigure}
\usepackage{epsfig}
\usepackage{color}
\usepackage{hyperref}
\usepackage{braket} 
\usepackage[T1]{fontenc}
\usepackage{amstext}
\usepackage{amssymb}
\usepackage{graphicx}
\usepackage{esint}
\usepackage{braket}
\usepackage{babel}
\usepackage{amsmath}
\usepackage{changes}
\usepackage{float}

\hypersetup{
colorlinks=true,
citecolor=blue,
linkcolor=red,
urlcolor=black
}


\newcommand{\e}{\textrm{e}}

\newcommand{\rr}{\mathbf{r}}

\newcommand{\asc}{a_{\textrm{\tiny 3D}}}
\newcommand{\aho}{l_\perp}
\newcommand{\atwod}{a_{\textrm{\tiny 2D}}}
\newcommand{\gTwoD}{g_{\textrm{\tiny 2D}}}
\newcommand{\gOneD}{g_{\textrm{\tiny 1D}}}

\newcommand{\aTwoD}{a_{\textrm{\tiny 2D}}}

\newcommand{\tildeg}{\tilde{g}_{\textrm{\tiny 2D}}}
\newcommand{\tildegOneD}{\tilde{g}_{\textrm{\tiny 1D}}}

\newcommand{\lettersection}[1]{\paragraph*{#1.---}}

\begin{document}

\title{
Leggett's bound and superfluidity in strongly interacting bosons
}

\author{Lorenzo Pizzino}
\affiliation{
DQMP, University of Geneva, 24 Quai Ernest-Ansermet, CH-1211 Geneva, Switzerland
}

\author{Haocong Pan}
\affiliation{
Institute of Quantum Electronics, School of Electronics, Peking University, Beijing 100871, China
}

\author{Thierry Giamarchi}
\affiliation{
DQMP, University of Geneva, 24 Quai Ernest-Ansermet, CH-1211 Geneva, Switzerland
}

\author{Hepeng Yao}
\affiliation{
Institute of Quantum Electronics, School of Electronics, Peking University, Beijing 100871, China
}

\date{\today}

\begin{abstract}
A density-based superfluid bound called Leggett’s bound has been proved to be a good estimator of the superfluid fraction for cold atomic gases in the mean-field regime. Here, we investigate the accuracy of such bound in the strongly interacting regime, where the mean-field approach fails. Combining quantum Monte Carlo, Gross-Pitaevskii equation and field-theory calculations, we demonstrate that the bound serves as a reliable estimator of the superfluid fraction for strongly interacting bosons at 2D-1D dimensional crossover at low temperatures. By further presenting two counterexamples where the bound predicts trivial results, we shed light on the conditions under which the Leggett’s bound serves as a good predictor.
\end{abstract}

\maketitle


Describing macroscopic quantum phenomena remains one of the central challenges in modern physics. Effects such as superconductivity, Bose-Einstein condensation (BEC), and superfluidity still have many aspects that remain theoretically and experimentally unexplored. A unifying feature of these phenomena is the emergence of collective behavior, encoded in a macroscopic wavefunction, that arises despite complex microscopic interactions~\cite{pitaevskii-book-2016}. Among them, superfluidity—first observed in 1938 \cite{Allen1938_Helium, Kapitza1938} as vanishing viscosity in Helium—offers a particularly striking example. The superfluid fraction $f_s$ quantifies the portion of a system that sustains dissipationless flow and is distinct from the condensate fraction, which measures macroscopic occupation of a single quantum state. While BEC and superfluidity often coexist, especially in weakly interacting systems, they can differ significantly in strongly interacting or low-dimensional settings~\cite{berezinskii1972,kosterlitz1973,hadzibabic-2Dgas-2011,giamarchi_book_1d}.

A significant theoretical advance in quantifying superfluidity was made by A. Leggett, who derived an upper bound for the superfluid fraction in systems with periodic boundary conditions \cite{leggett1970}.  Using a variational approach and considering the response to a phase twist, he showed that the superfluid fraction along a given direction $y$ satisfies the inequality
\begin{equation}
    f_s^y \leq f_{\uparrow,s}^y = \left\langle \frac{n_0}{\langle n(x,y) \rangle_x} \right\rangle_y^{-1}
    \label{eq:upper_bound_Leggett}
\end{equation}
with $n(x,y)$ the local density, $n_0$ the average total density, $x$ the other direction. The suppression of $f_s^y$ arises from the presence of low-density regions along the $x$-direction (\emph{transverse nodal surfaces}) which obstruct the current along $y$. A corresponding heuristic lower bound also exists \cite{leggett1998superfluid}. These bounds provide an intuitive density-based estimator of superfluidity, independent of detailed wavefunction knowledge. The upper bound has been analyzed in a series of papers and refined to study anisotropic SF \cite{Saslow1976}.

In cold atom experiments, differently from the BEC fraction which is directly measured from the time-of-flight technique, the detection of the superfluid fraction is challenging~\cite{davis1995, richard2003, stoferle-crossoverD-2004, gerbier2005,guo-crossoverD-2023}. Various complex methods have been proposed for such measurement, such as laser formed moving obstacle~\cite{desbuquois-stirring-2012}, observation of the sound velocity~\cite{christodoulou-2Dsuperfluid-2021,Dalibard2023_LeggettBound} and Josephson effect \cite{Biagioni_Josephson2024}. Thanks to the quantum gas microscope technique~\cite{bakr2009, weitenberg2011,Cheneau2012, greif2016,haller2015}, the spatial density distribution of cold atom systems becomes accessible. Therefore, if the Leggett's bound serves as a good estimator for the superfluid fraction, it provides an easy way to probe it. 

This question has been extensively investigated in the weakly interacting regime using the Gross-Pitaevskii equation (GPE), which offers access to both ground-state densities and phase structures \cite{Dalibard2023_LeggettBound,Spielman2023_LeggettBound}. By comparing the GPE based theoretical results, the Leggett's bound extracted from the experimental data of 
density distribution and the superfluid fraction measured from other methods, various groups have validated the bound as a useful proxy for $f_s$ in such regimes. 
Moreover, under the GPE mean-field scheme, the validity of the bound for other systems has been studied, such as disordered systems~\cite{Massignan2025_LeggettBound}, BEC mixtures~\cite{Massignan2025_LeggettBound}, triangular lattice~\cite{Beugnon_2025_tringularLegBound} and fermionic systems~\cite{stringari-SF-fermi-2024}.
This raises key open questions: Can the bound still yield quantitative predictions beyond mean-field regime, such as in strongly interacting systems? How does it behave at finite temperature or near dimensional crossover? Can the Leggett's bound capture more complex behavior and what is the limit of this bound?
Addressing these questions requires going beyond GPE and exploring the interplay of interaction, temperature, and dimensionality on the density-based criterion.

In this work, we test the accuracy of Leggett's bound, as a good estimator for the actual superfluid fraction, for strongly interacting bosons in the presence of an optical lattice driving the system from being two-dimensional (2D) to 1D. Combining quantum Monte Carlo (QMC) simulations with GPE solutions, we find the bound to be an excellent SF estimator even in the strong-interacting limit at low temperature, which is beyond the mean-field approximation. In the tight-binding regime, we use field-theory calculations to study the scaling of $f_s^y$ and understand in depth the success of the bound nearby the dimensional crossover point. Moreover, we show two examples where the bound cannot be used as a good estimator showcasing its limitation. We show that the bound accurately predicts the superfluid fraction as long as the mechanism of its variation is correlated to the density modulation of the system.


\lettersection{Model and numerical approach} 
\label{sec:model_and_methods}
The model we consider is described by the 2D Hamiltonian
\begin{equation}\label{eq:Hamiltonian}
    \hat{\mathcal{H}} = \sum_j \left[ -\frac{\hbar^2}{2m} \nabla^2_j + V(\hat{\rr}_j) \right] + \sum_{j<k} U(\hat{\rr}_j - \hat{\rr}_k)
\end{equation}
where the bosonic particles move under the effect of a periodic potential and the two-body repulsive interaction. We  denote with $\textbf{r}=(x,y)$ the spatial position of the particles, $U$ is the short-range interaction and $V(\textbf{r})=V_y\cos^2(ky)$ is the 1D lattice potential along y-direction with $k=\pi/a$ and $a$ is the lattice period.
In actual experiment, the interaction strength is determined by the 3D scattering length $\asc$ and the characteristic transverse confinement length $l_\perp$. They decide the 2D dimensionless coupling constant $\tildeg=m\gTwoD/\hbar^2$, see details in Refs.~\cite{Petrov2000a,petrov-2dscattering-2001,yao-crossD-2023}.

In this work, such Hamiltonian is exploited numerically by using Quantum Monte Carlo (QMC) \cite{Pizzino2025_crossD_finitesize, yao-crossD-2023} and Gross-Pitaevskii equation (GPE) \cite{gautier-2Dquasicrystal-2021}.  From the QMC side, we rely on path integral Monte Carlo in continuous space with worm algorithm implementations \cite{boninsegni-worm-short-2006, boninsegni-worm-long-2006}. Under the grand-canonical ensemble, we simulate the system at a given interaction strength $\tildeg$, temperature $T$ and chemical potential $\mu$.
By computing the winding number $W_i$ along the direction $i=x,y$ under periodic boundary conditions, we get the corresponding superfluid fraction $f_s^i$ defined as $f_s^i=m\langle W_i^2 \rangle/\beta \Omega\hbar^2$ with $\Omega$ the volume of the system and $\beta=1/k_BT$ the inverse temperature~\cite{ceperley-PIMC-1995}. We ensure that the selected low-temperature regime gives converged values of $f_s^{i,\text{QMC}}$, see Supplemental material for more details.
At the same time, by counting the closed worldlines in real space, we also compute the particle density distribution $n(x,y)$ from which we estimate the upper Leggett's bound $f^{i,\text{QMC}}_{\uparrow,s}$ by Eq.~(\ref{eq:upper_bound_Leggett}). As we use an unidirectional potential, the density factorizes $n(x,y)=n_0 n(y)$. For this potential, the \emph{strict} upper bound and \emph{putative} lower bound coincide, invalidating the second one as bound. In the remain we consider only the upper bound.

On the other hand, the GPE is expected to work in the limit of weak-interaction where a mean-field approach is valid. We also access the $f^{i,\text{GPE}}_{\uparrow,s}$ at zero temperature $k_BT=0$ by computing the density modulation $n(x,y)$ from the GPE solution similarly as \cite{ Dalibard2023_LeggettBound}. More precisely, the ground-state properties are governed by 
\begin{equation}
    \label{eq:GPE_equation}
    \mu \Psi = - \frac{\hbar^2}{2m} \nabla^2 \Psi  + V(\textbf{r}) \Psi  + gN |\Psi|^2 \Psi 
\end{equation}
with $\Psi=\Psi(\textbf{r})$ the Bose field and $g= (\hbar^2/m) \tilde{g}$ the coupling constant. The field is normalized as $\int d^2\textbf{r}|\Psi(\textbf{r})|^2=1$ and is computed by using the imaginary time evolution \cite{gautier-2Dquasicrystal-2021}. The (dimensionless) parameters are potential amplitude $V_y/E_r$ and the coupling constant $gn/E_r$ with $E_r$ the recoil energy.

\lettersection{Bound's quality beyond mean-field}

We start with exploring the Leggett's bound when interactions are strong such that the mean-field approach is not valid anymore. Indeed, for weak interactions the bound has been successfully exploited, especially by comparing cold atoms experiments with GPE simulations ~\cite{Dalibard2023_LeggettBound,Spielman2023_LeggettBound,Massignan2025_LeggettBound}.
The results show that the upper bound can be even used to quantitatively estimate the superfluid fraction for weak interaction regime. Here, by using ab initio QMC we are indeed able to go beyond the mean-field regime and test how accurate is the bond in both interaction regimes.

In Fig.~\ref{fig:comparison_fs_lowT}, we study the bound $f_{\uparrow,s}^y$ in the limit of low temperature as a function of the lattice potential $V_y$ for both weak and strong interactions. We take the temperature $k_BT\simeq 0.005E_r$ where the system is below quantum degeneracy in both 2D and 1D limits. We use QMC and GPE algorithm to extrapolate the density $n(x,y)$ which ranges from being totally flat for $V_y=0$ to effectively being sliced in 1D chains for very large $V_y$. Such quantity is then used to compute the upper Leggett's bound $f^{y,\text{GPE}}_{\uparrow,s}$ and $f^{y,\text{QMC}}_{\uparrow,s}$ from Eq.~\eqref{eq:upper_bound_Leggett}. Both estimators are compared in Fig.~\ref{fig:comparison_fs_lowT} with the exact $f_s^y$ computed from the QMC winding number. In both weak (Fig.~\ref{fig:comparison_fs_lowT}(a)) and strong (Fig.~\ref{fig:comparison_fs_lowT}(b)) interaction regimes, we observe that the upper bound computed from the QMC particle density not only works as a bound but also seems to be very close to the actual value of $f_s^y$ for the full range of $V_y$ we considered. To our knowledge, this is the first demonstration that the bound also applies as a good estimator in the strong-interacting limit where a mean-field approach would not be suitable. Clearly, for a finite $V_y$ at zero temperature we have a density modulation along the y-direction with a consequent non-zero component of normal fluid and a lower SF fraction $f_s^y$. This is reflected in the discrepancy of the  $f^{y,\text{GPE}}_{\uparrow,s}$ with the QMC result, see solid red line in Fig.~\ref{fig:comparison_fs_lowT}(b). 
Moreover, for weak interactions, we find the GPE results are also in excellent agreement with QMC as expected, while it overestimates the superfluid fraction in the strongly interaction limit.

\begin{figure}[t!]
    \centering
    \includegraphics[width = 1 \columnwidth]{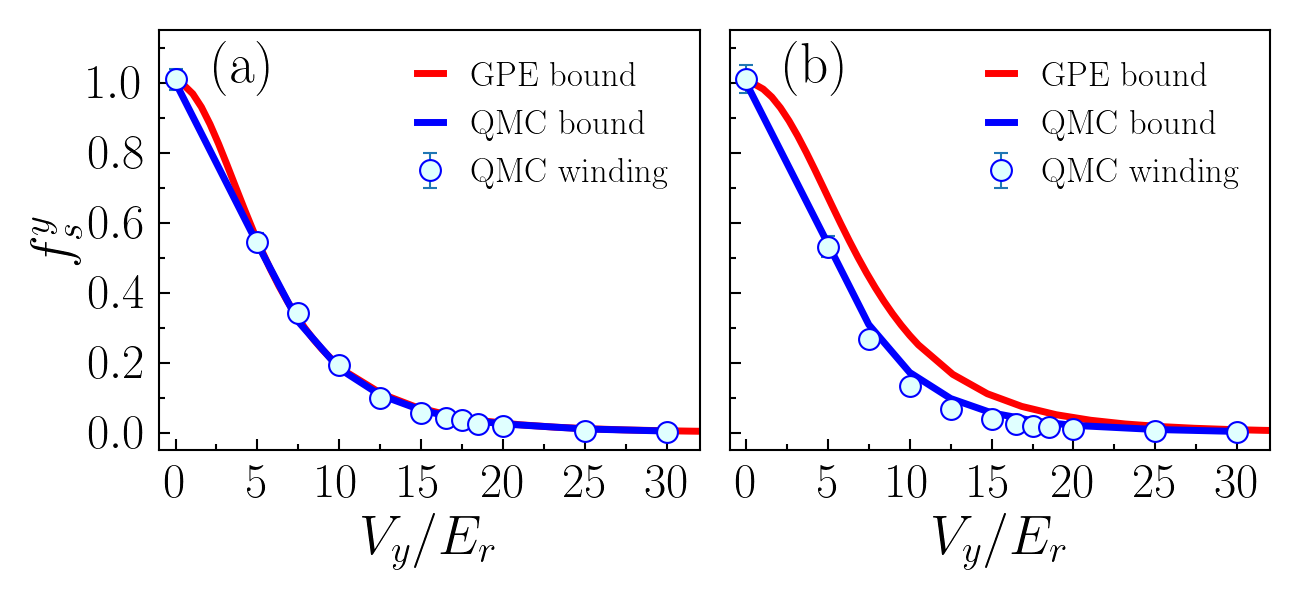}
    \caption{Comparison between the exact superfluid fraction $f_s^y$ via QMC winding number (blue circles) and the Leggett's bound $f_{\uparrow,s}^y$ (solid lines). We use QMC at low temperature $k_BT\simeq 0.005E_r$ and GPE density profiles at zero temperature $k_BT=0$ (blue and red solid lines, respectively) for varying lattice depths $V_y$. Panels (a) and (b) correspond to weak ($\tildeg\simeq0.018$) and strong ($\tildeg\simeq1.364$) interaction regimes, respectively. The QMC bound shows excellent agreement with the exact $f_s^y$ in both cases. As expected, the GPE bound is more accurate in the weak interaction case.
    \label{fig:comparison_fs_lowT}}
\end{figure}


\lettersection{Scaling for large $V_y$}
\label{sec:bound_interaction_main}

To further understand the accuracy of Leggett's bound in the strongly interacting regime in depth, we study the limit of large $V_y$ where the system is well described by a tight-binding model of 1D chains. In this limit, we investigate the system under the Tomonaga-Luttinger liquid (TLL) theory ~\cite{giamarchi_book_1d}. Compared to the Hamiltonian Eq.~\eqref{eq:Hamiltonian}, which is defined in the continuum, here we account the effect of the transverse coupling in the limit of large $V_y$ \cite{yao-crossD-2023} by considering instead the hopping term $t_y/E_r=\frac{4}{\sqrt{\pi}}(V_y/E_r)^{3/4}e^{-2\sqrt{V_y/E_r}}$ \cite{bloch-review-2008} between nearest neighbors chains along $y$ direction using a variational approach. In this regime, the strong potential confines particles into weakly-coupled tubes, effectively realizing a quasi-1D array. 

Following the same footsteps of Ref.~\cite{cazalilla-coupled1D-2006} (see details in Supplemental material), we compute the free energy within the self-consistent harmonic approximation (SCHA) and use the definition of the superfluid fraction \cite{fisher1973} to find
\begin{equation}
    f^{y,\text{SCHA}}_s \sim \frac{d^2 \mathcal{F}_\text{SCHA}}{d \Phi ^2} \Bigg \vert_{\Phi \rightarrow 0}  \sim \left ( \frac{t_y}{E_r}\right)^{\nu(K)}
    \label{eq:f^y_s_SCHA}
\end{equation}
with the scaling exponent being $\nu(K)=\frac{4K}{4K-1}$ and $K$ the Luttinger parameter ($K=1$ in the limit of hard-core bosons and $K=10$ for soft-core bosons) and $\Phi$ a small enough flux threading the system: the response of the system to such flux describes its superfluid nature. Given the scaling, the superfluid fraction becomes more sensitive to the interchain coupling $t_y/E_r$ as interactions weaken (increasing $K$). Moreover, such scaling $\nu$ is not affected by finite size \cite{yao-crossD-2023, Pizzino2025_crossD_finitesize}.

In Fig.~\ref{fig:fs_LowT_vs_analytical_small_tperp}, we test the SCHA scaling with QMC simulations, which results to be more exact than GPE for the values of interactions we consider, already presented in Fig.~\ref{fig:comparison_fs_lowT} but in log-log scale. We consider data that correspond to $f_s^y > 0.005$ and check the scaling exponent according to \cite{yao-crossD-2023,Pizzino2025_crossD_finitesize} by fitting in the range $V_y \in [15, 20]E_r$. Starting from the weak-interaction limit in Fig.~\ref{fig:fs_LowT_vs_analytical_small_tperp}(a), we perform a linear fit which gives a scaling exponent of $\nu^{\text{QMC}}_{\text{weak}} = 1.06 \pm 0.09$ which is in excellent agreement with $\nu(K=10) \simeq 1.03$. On the other hand, the bound gives a scaling of $\nu^{\text{QMC bound}}_{\text{weak}}= 0.96 \pm 0.01$. We thus see that although the two results agree very well quantitatively as shown in Fig.~\ref{fig:fs_LowT_vs_analytical_small_tperp}(a), the bound and the full calculation indicate different scalings as a function of $t_y$. The bound will thus become more and more inaccurate as $t_y \to 0$. This is due to the quantum effects coming from the interactions that cannot be captured by a density description alone. However for weak couplings the exponents are so close that it would require abysmally small $t_y$ to have a sizeable difference. The bound is thus a remarkable estimator in practice for all reasonable values of $t_y$ as shown in
Fig.~\ref{fig:fs_LowT_vs_analytical_small_tperp}(a)

This trend becomes more obvious for the strongly interacting case shown in Fig.~\ref{fig:fs_LowT_vs_analytical_small_tperp}(b). The winding number fit gives a scaling exponent of $\nu^\text{QMC}_\text{strong} = 1.33 \pm 0.07$ which is again in excellent agreement with the field-theory prediction $\nu(K=1)=4/3$. At the same time, compared to the bound, the scaling exponent we find is $\nu^{\text{QMC bound}}_{\text{strong}} = 1.00 \pm 0.03$. Therefore, for strong interactions the bound is unable to predict the correct scaling, contrarily to the case of weak interactions, even though the discrepancy between the QMC and bound is not dramatic. This result shows that the density itself is insufficient to infer on the effect of interactions on the system. Furthermore, as we move towards 1D geometry, the effect of quantum fluctuations are enhanced and even for an \emph{almost} constant density the bound predicts trivial results. Indeed, the ratio between the winding number and bound suggests that the bound becomes increasingly more incorrect for small $t_y$.

\begin{figure}[t!]
         \centering
         \includegraphics[width = 1\columnwidth]{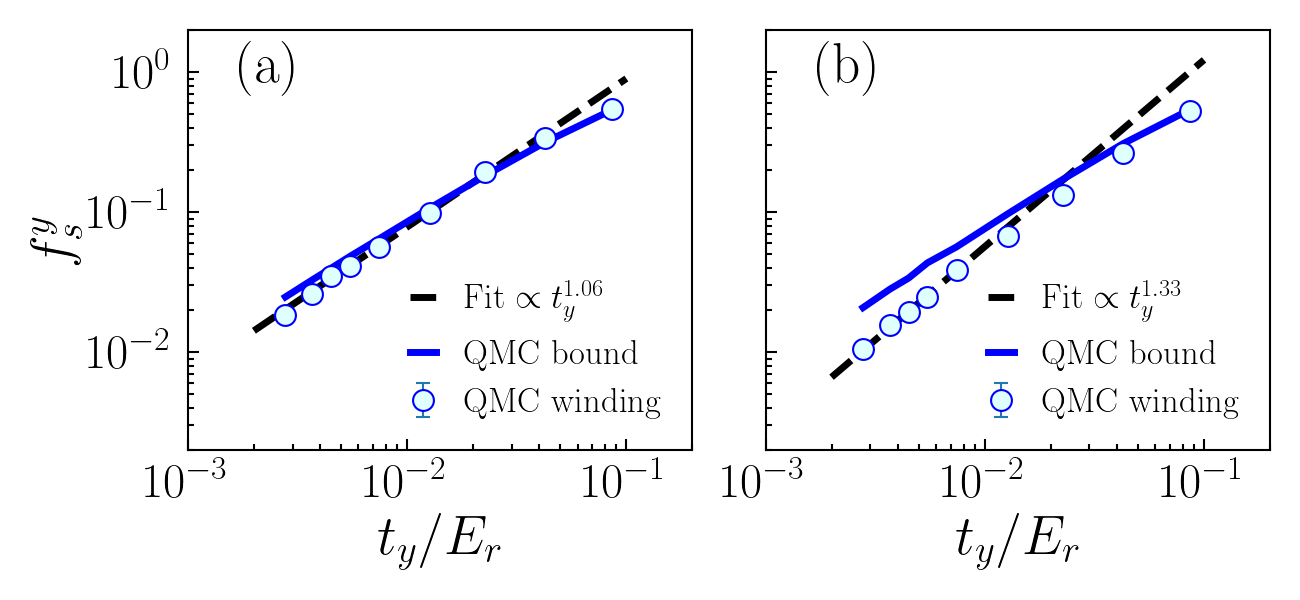}
    \caption{ \label{fig:fs_LowT_vs_analytical_small_tperp} Scaling analysis of $f_s^y$ for small $t_y$ from QMC data. Panels (a) and (b) correspond to weak ($\tildeg\simeq0.018$) and strong ($\tildeg\simeq1.364$) interaction regimes, respectively, for a system size of $L_x=25a, L_y=10a$ and $k_B T\simeq 0.005 E_r$. The black dashed lines are linear fits for small $t_y$ of the exact SF fraction computed with QMC winding number (light blue circles) and the blue solid line is the QMC bound. We find the scaling exponents to be in excellent agreement with the SCHA predictions $\nu(K)=\frac{4K}{4K-1}$. For weak interactions ($K=10$) we have $\nu^{\text{QMC}}_\text{weak}=1.06 \pm 0.09$ and for strong interactions ($K=1$) we find $\nu^{\text{QMC}}_\text{strong}=1.33 \pm 0.07$. Fitting the QMC bounds, we have $\nu^{\text{QMC bound}}_{\text{weak}} = 0.96 \pm 0.01$ and $\nu^{\text{QMC bound}}_{\text{strong}} = 1.00 \pm 0.03$. } 
\end{figure}


\lettersection{Two special cases where the bound strongly overestimates the superfluid fraction}
\label{sec:special_case_complementary}

So far we have shown how the Leggett's bound behaves as a good estimator for the superfluid fraction of low temperature systems as a function of interactions. Here, we want to further demonstrate that it is not always the case by presenting two counter examples. Firstly, we explore the dimensional crossover at intermediate temperature regime with $\tildeg=1.36$ and $k_BT=0.09E_r$. At this temperature, the system is a 2D superfluid, but becomes a normal fluid when it enters 1D regime (increasing $V_y$). Therefore, the longitudinal superfluid fraction $f_s^x$ will decrease as a function of $V_y$ according to Ref.~\cite{yao-crossD-2023}. 
In Fig.~\ref{fig:inaccurate_Leggett_1D}(a1), we compute the $f_s^x$ from the winding number as $V_y$ (light blue balls). Clearly, we see that $f_s^x$ decreases with $V_y$ as expected. However, by plotting the density profile along $x$-direction in Fig.~\ref{fig:inaccurate_Leggett_1D}(a2), we find the absence of density modulation which leads to a constant upper bound, see black solid line in Fig.~\ref{fig:inaccurate_Leggett_1D}(a1). This is due to the fact that the decrease of $f_s^x$ is caused by the joint effect of thermal and quantum fluctuations along the longitudinal directions, which does not necessarily reflect on the density profile. Therefore, the computed bound cannot encode such information and is unable to follow the variation of superfluid fraction.

As a second example, we start off with a large potential $V_y$ by forcing the system to be in the 1D limit (large $V_y$) at low temperature. We now have to recall that for purely 1D systems in strong interaction regime, an arbitrarily weak periodic lattice potential can localize particle and destroy even at zero temperature a finite SF fraction, known as the pinning Mott transition~\cite{haller-pinning-2010, boeris-1dshallow-lattice-2016, yao-boseglass-2020, DErrico_2024_delocalization}.
From the bound’s perspective, such a potential only mildly affects the density modulation and therefore can potentially lead the bound to give a trivial result. We show in Fig.~\ref{fig:inaccurate_Leggett_1D}(b1) the Mott transition in such a shallow lattice case. Here, the potential along x-direction writes $V(x)=V_x\cos^2(kx)$ with $k=\pi/a$. More precisely, we take the parameters $\tildegOneD=7$ and $V_x=2E_r$, with a temperature of $k_BT\simeq0.004E_r$ and plot $f_s^x$ as a function of the chemical potential $\mu$ around the Mott lobe $na=1$. The transition point is located in the window $\mu_{c1}=1.38E_r$ and $\mu_{c2}=1.51E_r$, and within the range $\mu_{c1}<\mu<\mu_{c2}$ the superfluid fraction $f_s^x$ vanishes. However, such a small oscillating potential $V_x = 2E_r$ only causes a weak modulation on the corresponding density profile. In Fig.~\ref{fig:inaccurate_Leggett_1D}(b2), we show one example of the density profile for $\mu=1.45E_r$ in the center of the Mott lobe. Clearly, it only oscillates between $na\simeq0.48$ and $na\simeq1.65$ without touching zero. As a consequence, the corresponding Leggett's bound gives a large value $f_{\uparrow,s}^x\simeq0.827$. 
In fact, in the full range of $\mu$ we considered, the Leggett's bound is not sensitive to $\mu$ and shows almost a constant, see blue solid line in Fig.~\ref{fig:inaccurate_Leggett_1D}(b1). Hence, the density modulation is one of the ingredients to suppress SF but not the only one: if more subtle mechanisms are involved, then the upper bound usually gives \emph{trivial} results as shown in this section.

\begin{figure}[t!]
    \centering
    \includegraphics[width = 1\columnwidth]{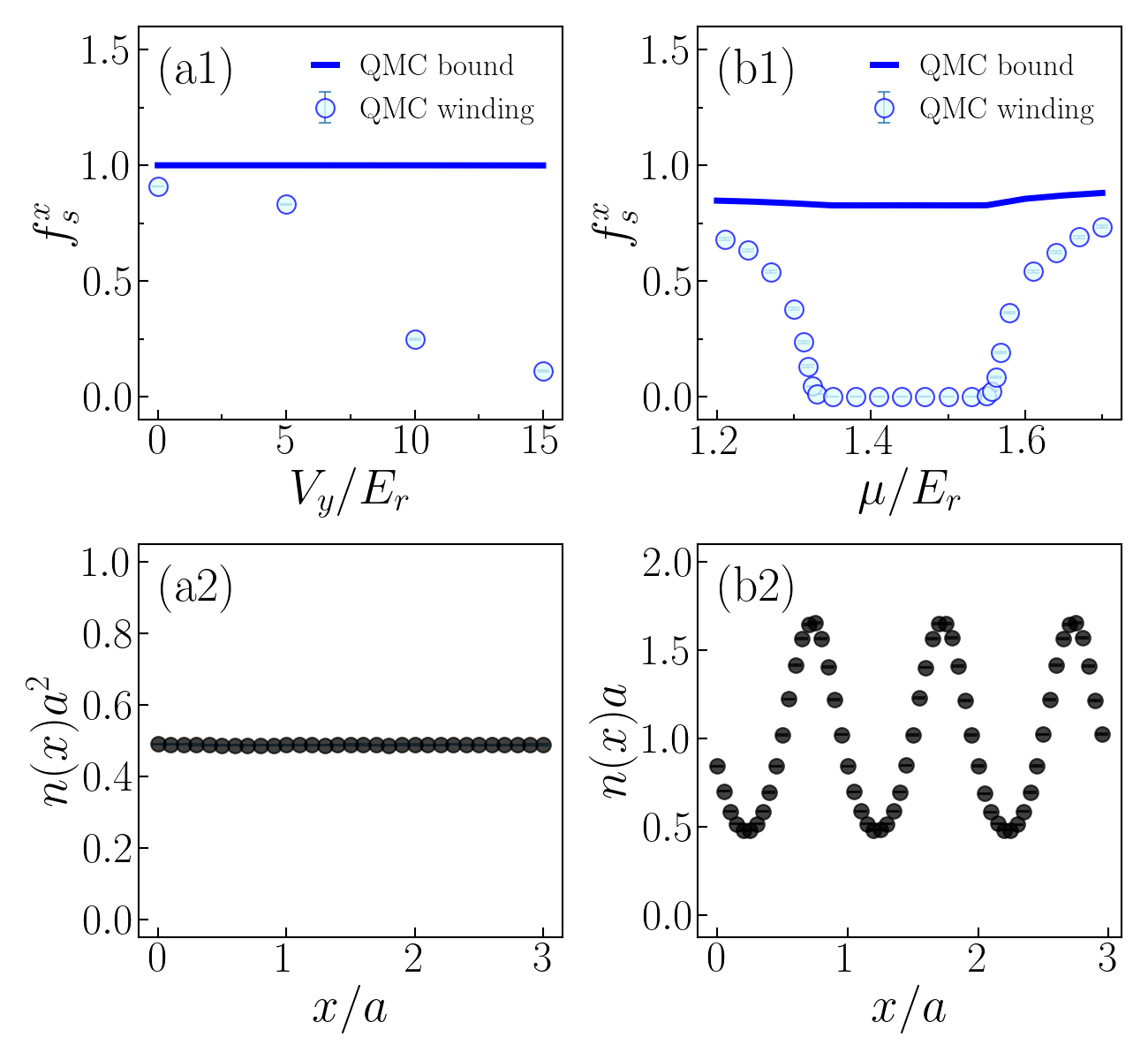}
    \caption{\label{fig:inaccurate_Leggett_1D} Two cases where the bound is no more a quantitative SF estimator (solid line). In panel (a1) we study the longitudinal SF fraction $f_s^x$ at intermediate temperature $k_BT=0.09E_r$ and strong interactions $\tildeg\simeq1.364$ with system size of $L_x=L_y=12a$. (a2) shows the corresponded density profile $n(x)$ for the case $V_y=10E_r$. It leads to the constant bound (blue solid curve in (a1) which is not compatible with the $f_s^x$ (light blue balls). In panel (b1) we consider large enough $V_y$ such that the system is strictly 1D. We further consider small $V_x=2E_r$ in the Tonks-Girardeau limit , with the 1D dimensionless coupling constant $\tildegOneD=7$ and temperature $k_BT\simeq0.004E_r$ and system sizes $L_x=50a$. The system shows a transition from 1D superfluid state to 1D Mott insulator state by plotting the superfluid fraction $f_s^x$ as a function of the chemical potential $\mu$. We show the results from both the QMC winding (light blue balls) and Leggett's bound (blue solid curve). In panel (b2) we present one example of the density profile $n(x)$ at $\mu=1.45E_r$.}
\end{figure}


\lettersection{Conclusion and experimental observability}

In this paper, we shed light on the conditions under which the Leggett's bound serves as a good estimator for superfluid fraction of strongly interacting bosons. For bosonic systems at 2D-1D dimensional crossover and low temperature, we find the bound is an accurate estimator of the transverse superfluidity for both strong and weak interaction regimes. While the weak interaction limit was already widely studied, we get further insights from the SCHA scaling on the strong interaction limit. The results confirm that the bound scaling agrees with QMC and SCHA in the limit of weak-interactions, while the discrepancy becomes larger for strong-interactions. Therefore, we conclude that the bound serves as a quantitative superfluid estimator regardless of the value of interactions but is not able to fully predict the effect of interactions in the limit of quasi-1D. We further show two additional cases where the bound \emph{fails} to be a good superfluid estimator, namely the dimensional crossover at higher temperature and the pinning Mott transition in 1D Tonks-Girardeau regime. In both cases, the density is weakly modulated and we find $f_{\uparrow,s}^x \sim 1$. We conclude that the bound is an extremely powerful tool to study the superfluid fraction when the superfluid suppression mechanism is linked to density modulation. Effects like thermal fluctuations and interferences can also cause changes on superfluid fraction while making less influence on the density, which make the bound less powerful. We recall that the examples we presented are intimately linked with the peculiar case of 1D quantum systems. Such a scenario is more difficult to occur in 2D or 3D where, unless we have finite range interactions or disorder \cite{Massignan2025_LeggettBound}, the localization (density modulations) can be indeed externally imposed with non-trivial lattices \cite{rabec2025_bound} or self-induced (supersolids).

Concerning nowadays cold atom experiments, we stress out that such tool is very useful, especially for situations where the density profile is easy to obtain. For instance, single-site resolution imaging via quantum gas microscopes~\cite{bakr2009, weitenberg2011,Cheneau2012, greif2016,haller2015} provides direct access to the density profile, making Leggett's bound particularly appealing as an indirect probe of superfluidity. Moreover, on the theoretical side, it would be interesting to study the validity of the bound for strongly interacting bosons in more complex potentials, such as speckle or quasiperiodic potentials. Also, it is interesting to explore the bound in case of fermionic particles which could present a different physical scenario compared to the bosonic case. 

\begin{acknowledgments}
We thank Pietro Massignan, Jean Dalibard, Sandro Stringari and Grigory Astrakharchik for fruitful discussions and useful comments. We thank Wayne Saslow for useful comments and making us aware of relevant papers on the subject of Leggett's bounds. T.G. thanks the Institut Henri Poincaré (UAR 839 CNRS-Sorbonne Université). This work is supported by the Swiss National Science Foundation under grant number 200020-219400 and The Fundamental Research Funds for the Central Universities, Peking University. Numerical calculations make use of the ALPS
scheduler library and statistical analysis tools \cite{troyer1998, ALPS2007, Bauer-JSTAT-2011}.
\end{acknowledgments}

\bibliography{biblio_final}


 \renewcommand{\theequation}{S\arabic{equation}}
 \setcounter{equation}{0}
 \renewcommand{\thefigure}{S\arabic{figure}}
 \setcounter{figure}{0}
 \renewcommand{\thesection}{S\arabic{section}}
 \setcounter{section}{0}
 \onecolumngrid  

\newpage

\maketitle


 \renewcommand{\theequation}{S\arabic{equation}}
 \setcounter{equation}{0}
 \renewcommand{\thefigure}{S\arabic{figure}}
 \setcounter{figure}{0}
 \renewcommand{\thesection}{S\arabic{section}}
 \setcounter{section}{0}
 \onecolumngrid

 {\center \bf \large Supplemental Material for \\}
 {\center \bf \large Leggett's bound and superfluidity in strongly-interacting bosons \\ \vspace*{1.cm}
 }

In this supplemental material, we provide complementary details regarding the analytic and numerical results presented in the manuscript.

\section{S1. Superfluid fraction scaling from variational method}

In this section, we show how to find the superfluid fraction along the y-direction in the limit of 1D coupled chains by using a field-theory approach. We start off by the single-particle operator which in the bosonization dictionary reads $\psi(x) = \sqrt{A_B \rho_0}\, e^{i\theta(x)}$ with $A_B$ a non-universal model-dependent pre-factor and $\rho_0$ the density of particles. Here, $\theta$ is the field corresponding to the phase of the bosonic particles. The action of the coupled chains is
\begin{equation}
    \mathcal{S} = \mathcal{S}_0-A_B\rho_0 t_y \sum_{\langle i,j \rangle} \int dx d\tau \, \cos \left ( \theta_i(x,\tau)-\theta_j(x,\tau)\right )
\end{equation}
with $S_0$ being the action of the single-chain and the second term generated by the very weak transverse coupling. In our notation, $x$ is the (longitudinal) spatial coordinate and $\tau$ the imaginary time coordinate while $\langle i,j\rangle$ denote n.n. chains indices. 

In order to compute $f^y_s$, we perform a gauge transformation along the $y$-direction such that $\theta_j(x,\tau) \rightarrow \theta_j(x,\tau) + \Phi j$ and therefore we modify the argument of the cosine while $\mathcal{S}_0$ remains unchanged. We denote with $\Phi$ a small flux threading the system. Next, in the limit of small temperature (small fluctuations) we consider the contribution coming from small oscillations of the cosine which results in having an effective action of the form
\begin{equation}
    \begin{aligned}
        \mathcal{S} &= \mathcal{S}_0-A_B\rho_0 \tilde{t}_y \sum_{\langle i,j \rangle} \int dx d\tau \, \left[1- \frac{1}{2}\Big((\theta_i(x,\tau)-\theta_j(x,\tau))^2 + \Phi^2 \Big)\right] \\
        &= \mathcal{S}_0 + \frac{A_B\rho_0 \tilde{t}_y}{2}\sum_{\langle i,j \rangle} \int dx d\tau \,  (\theta_i(x,\tau)-\theta_j(x,\tau))^2 + \frac{A_B\rho_0 \tilde{t}_y}{2}\frac{L_y z}{2} L_x\beta\,\Phi^2  + \text{const}\\
    \end{aligned}
\end{equation}
where $\tilde{t}_y$ has to be found self-consistently and has a known scaling of $\tilde{t}_y \sim t_y^{\frac{4K}{4K-1}}$ \cite{cazalilla-coupled1D-2006}. It's of our interest to keep track of the $\Phi^2$ term. The factor $\frac{L_yz}{2}$ comes from counting the number of n.n. along the y-direction, $\beta$ is the inverse temperature and $L_x$ is the system size along the longitudinal direction. From the action, we have access to the partition function which allows us to compute the free energy as
\begin{equation}
    \begin{aligned}
        \mathcal{F} = -k_BT \ln \left [ e^{-\frac{1}{4}A_Bz\rho_0 \tilde{t}_yL_y L_x\beta\,\Phi^2} \int \mathcal{D}\theta^* \int \mathcal{D}\theta^{}e^{-\frac{1}{2\beta L L_y} \sum_{\textbf{Q}}\left [\dots\right] \theta^*_\textbf{Q} \theta^{}_\textbf{Q}} \right] \sim \frac{1}{4}A_Bz\rho_0 \tilde{t}_yL_y L_x\,\Phi^2
    \end{aligned}
\end{equation}
and the superfluid fraction reads
\begin{equation}
    f_s^y \sim \frac{d^2\mathcal{F}}{d\Phi^2} \sim \tilde{t}_y\, \sim t_y^{\frac{4K}{4K-1}}
\end{equation}
which gives the power-law scaling as a function of the Luttinger parameter $K$.

\section{S2. Quantum Monte Carlo calculations}
\label{sec:QMC}

The calculations of superfluid fraction and its upper bound in the main paper highly rely on one type of quantum Monte Carlo (QMC) calculations, namely path integral Monte Carlo implemented with worm algorithm~\cite{boninsegni-worm-short-2006,boninsegni-worm-long-2006}.
Thanks to the open worldline configurations in worm algorithm, we work under the grand canonical ensemble.
Given temperature $T$, 2D scattering length $\atwod$ and chemical potential $\mu$, relevant observables that we are interested in are computable.
In this section, we provide more details about the QMC calculations.

\subsection{Computation of superfluid fraction}

We further compute the superfluid fraction $f_s^x$ or $f_s^y$ and their upper bounds $f_{\uparrow,s}^x$ or $f_{\uparrow,s}^y$ from two different methods, respectively.
On one hand, we perform the winding method under periodic boundary conditions. More specifically, the definition of superfluid mass along certain direction $i$ is
\begin{equation}
M_s^i=M-\frac{\partial\langle\hat{P}_i\rangle}{\partial v},
\end{equation}
where $v$ is the velocity with respect to the reference frame, $M$ the mass of the system, and $\hat{P}_i$ the momentum operator of the system along certain direction $i$, where $i=x,y$.
Then, the superfluid density can be written as $n_s^i=M_s^i/mL_xL_y$, where $m$ is the mass of a single particle.
Recalling that the thermodynamic averages of an observable $\hat{A}$ can be estimated by
\begin{equation}\label{eq:app:avA}
\langle\hat{A}\rangle = \frac{\textrm{Tr}\left[\e^{-\beta(\mathcal{H}-\mu\hat{N})}\hat{A}\right]}{\textrm{Tr}\left[\e^{-\beta(\mathcal{H}-\mu\hat{N})}\right]},
\end{equation}
where $\mathcal{H}$ is the Hamiltonian, $\hat{N}$ the number of particles operator, $\beta=1/k_BT$ the energy scale of inverse temperature, and $\textrm{Tr}$ the trace operator~\cite{ceperley-PIMC-1995}. By inserting $\hat{A}=\hat{P}_i$ and $\mathcal{H}=\hat{H}_v=\hat{H}-\mathbf{v}\cdot\hat{\mathbf{P}}$, we have
\begin{equation}\label{eq:med_step_fs}
n_s^i=n_0-\frac{m}{L_xL_y}\int_0^\beta d\tau \, \langle \hat{P}_i(\tau)\hat{P}_i(0) \rangle,
\end{equation}
where $n_0$ is the average total density.

To calculate the correlator $\langle \hat{P}_i(\tau)\hat{P}_i(0) \rangle$ in the integral, we replace the integral $\int_0^\beta$ by the discrete sum $\epsilon\sum_{j=0}^{J-1}$ by slicing the imaginary time space into $J$ small time step $\epsilon$, such that $\epsilon=\beta/J$.
Then, we take $\epsilon\to0$ and obtain
\begin{equation}\label{eq:fs-QMC}
n_s^i=\frac{1}{\beta L_xL_y}\frac{m}{\hbar^2}\langle W_i^2 \rangle, \ i=x,y,
\end{equation}
with $W_i$ the winding number along $i$ direction under the periodical boundary conditions. Finally, we can further obtain the superfluid fraction by $f_s^i=n_s^i/n_0, \ i=x,y$, with
\begin{equation}\label{eq:average_density}
n_0=\frac{1}{L_xL_y}\langle\hat{N}\rangle,
\end{equation}
where the particle number $\langle\hat{N}\rangle$ can be directly obtained by counting the number of worldlines.

On the other hand, the upper bound of the superfluid fraction is linked to the density profile by a concise formula, namely Eq.~(1) of the main text\cite{leggett1970, Dalibard2023_LeggettBound}.
Since the local density is diagonal in the position representation, and using the translational invariance in the imaginary time of the path integral configurations, we can calculate the density distribution $n(\mathbf{r})$ in the simulation as
\begin{equation}\label{eq:density_profile}
n(\mathbf{r})=\frac{1}{J}\left\langle \sum_{j=1}^{J-1}\sum_{i=0}^N\delta(\mathbf{r}-\mathbf{r}_i^j) \right\rangle,
\end{equation}
where the superscript of $\mathbf{r}$ represents the number of steps in the imaginary time evolution and the subscript represents the label of a certain particle. With the computed $n(\mathbf{r})$, we can then further compute the upper bound $f_s^i$ according to Eq.~(1) of the main text.

\subsection{Depiction of interaction}

In the QMC calculations, we use the Trotter-Suzuki approximation for estimating the short-time propagator, and only consider two-body interaction under pair-product approximation, where a generalized 2D interaction propagator that can work for any interaction regime is generated, similarly as Ref.~\cite{gautier-2Dquasicrystal-2021}. 

When we use a strong harmonic confinement $\omega_\perp$ on the transverse direction to generate a 2D gas, the 2D scattering length $\atwod$ can be expressed as a function of the 3D scattering length $\asc$ and characteristic transverse length $\aho=\sqrt{\hbar/m\omega_\perp}$~\cite{Petrov2000a,petrov-2dscattering-2001}, which writes
$ \atwod \simeq 2.092\aho \exp\left(-\sqrt{{\pi}/{2}} {\aho}/{\asc}\right)$.
For 2D or 1D cases, we can link the scattering length with 2D coupling constant $\gTwoD$ or 1D coupling constant $\gOneD$, which describe the pseudo-potential strength~\cite{bloch-review-2008}.
The 2D coupling constant $\gTwoD$ can be written as~\cite{Petrov2000a,petrov-2dscattering-2001}
\begin{equation}\label{eq:g2d}
\tildeg \simeq \frac{4\pi}{2\ln(a/\atwod) + \ln\left(\Lambda E_r/\mu\right)},
\end{equation}
with $\Lambda \simeq 2.092^2/\pi^3 \simeq 0.141$, $E_r = \pi^2\hbar^2/2ma^2$ the recoil energy, and $\tildeg=m\gTwoD/\hbar^2$. The 1D coupling constant $\gOneD$ can be written as~\cite{olshanii-1Dscattering-1998,bloch-review-2008}.
\begin{equation}\label{eq:g1d}
\tildegOneD= \frac{2 \asc}{\aho^2} \left(1-\frac{1.036\asc}{\aho}\right)^{-1},
\end{equation}
with $\tildegOneD=m\gOneD/\hbar^2$. The definition we used here is the same as the one in Ref.~\cite{yao-crossD-2023}.

In this paper, we consider $\aTwoD=10^{-150}a$ for 2D weak interactions and $\aTwoD=0.01a$ for 2D strong interactions, where $a$ is the lattice period, and particle density $n = N/L_xL_y=0.5a^{-2}$, which is satisfied by choosing proper chemical potential $\mu$.
Since the condition $a/\aTwoD \gg \Lambda E_r/\mu$ is satisfied in both 2D weak and strong interactions, the 2D dimensionless coupling constant can be reduced to $\tildeg\simeq 2\pi/\ln(a/\aTwoD)$, which gives $\tildeg\simeq0.018$ for 2D weak interactions and $\tildeg\simeq1.364$ for 2D strong interactions.
Notably, the criteria of strongly-interacting bosons is $\gamma_{\textrm{\tiny 2D}}=\tildeg\gtrsim1$, see Refs.~
~\cite{bloch-review-2008,hadzibabic-2Dgas-2011,cazalilla_review_bosons}.
For the strictly-1D calculation in Fig. 3 (b) of the main text, we consider the 1D dimensionless coupling constant $\tildegOneD=7$. For particle density around $na=1$, it leads to the Lieb-Liniger parameter $\gamma_{\textrm{\tiny 1D}}=\tildegOneD/na=7$, which also satistifies the criteria of strong interaction regime $\gamma_{\textrm{\tiny 1D}}\gg 1$.

\subsection{Numerical parameters and validity of approximations}

Here, we provide the numerical parameters we choose, which ensure the validity of approximations and minimizes the numerical errors.

For the imaginary time space discretization, we use small imaginary time steps $\epsilon=0.01-0.02E_r^{-1}$ and $\epsilon=0.05E_r^{-1}$ for the 2D and 1D cases, respectively.
Both of them enable us to use the Trotter-Suzuki approximation and the pair-product approximation by meeting the criteria $\epsilon\ll\beta$ and $\sigma=\sqrt{\hbar^2\epsilon/m}\ll a$, where $\sigma$ is the corresponding standard deviation of the free particle propagator.

Besides, a sufficiently large number of warm-up steps and sampling iterations are needed to guarantee the adequacy of the Monte Carlo statistics.
For the 2D cases, we take $10^8$ warm-up steps for most of the data points, with $\sim 10^7$ iterations for both strong and weak interactions. For the 1D cases, we take $10^6$ each for warm-up steps and iterations.
Finally, we have make sure that a larger number of warm-up steps and sampling iterations will not change any physics presented in the main paper.

\subsection{Convergence of superfluid fraction for small temperatures}

In the main text, for all the QMC calculations in Figs.~1 and ~2, we always use the temperature $k_BT=0.005E_r$. We assume this temperature is low enough such that there is no finite-temperature effect in the obtained results. Here, we examine this point by computing the superfluid fraction at various temperatures for the strong interaction case, where a deviation between the Leggett's bound and the winding superfluid fraction is observed.

\begin{figure}[t!]
    \centering
    \includegraphics[width = 0.6 \columnwidth]{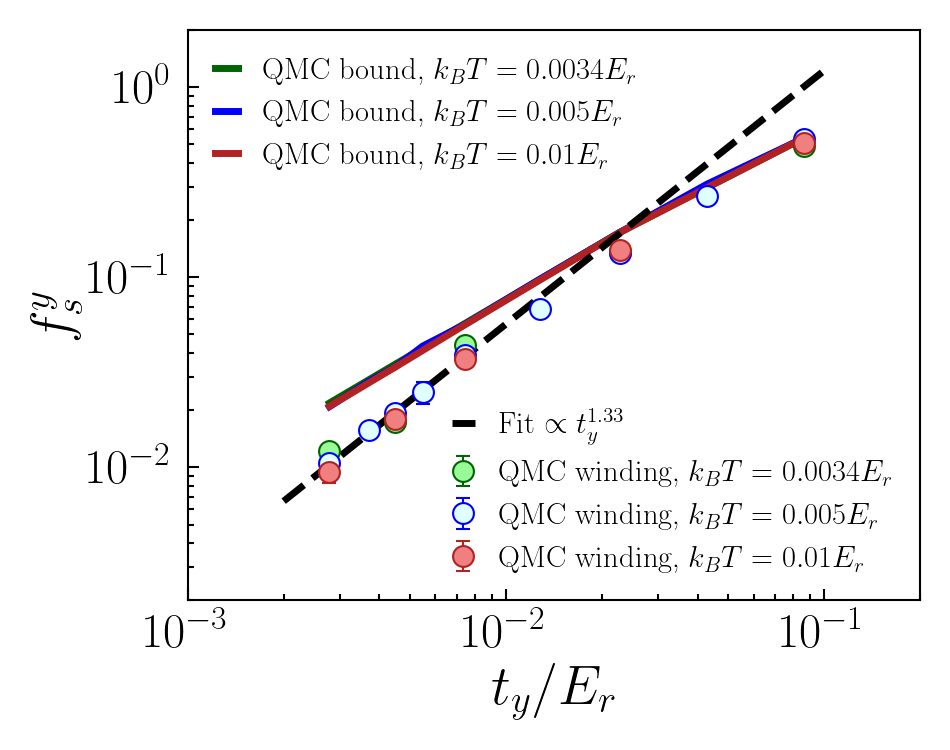}
    \caption{Comparison of QMC calculations for different temperatures $k_BT=0.0034E_r$ (green), $k_BT=0.005E_r$ (blue) and $k_BT=0.01E_r$ (red) in strong  interaction regime $\tildeg\simeq1.364$, with the system size $L_x=25a, L_y=10a$.
    The solid lines are Leggett's bounds, with the linear fit in log-log scale of the superfluid fraction $f_s^y$ (black dashed line) computed via QMC winding number at $k_BT=0.005E_r$ in the range $V_y\in[15,20]E_r$ (blue circles).
    }
    \label{fig:supplemental_temp_check}
\end{figure}

As shown in Fig.~\ref{fig:supplemental_temp_check}, on top of the results in Fig.~2 (b) of the main text, we further examine the cases of a lower temperature $k_BT=0.0034E_r$ and a higher temperature $k_BT=0.01E_r$ under strong interaction at $V_y/E_r=5,10,15,17.5,20$, with $t_y/E_r=\frac{4}{\sqrt{\pi}}(V_y/E_r)^{3/4}e^{-2\sqrt{V_y/E_r}}$ \cite{bloch-review-2008}.
Clearly, there is no obvious difference in the Leggett's bounds at the three temperatures. The superfluid fractions $f_s^y$ computed via QMC winding number are also consistent with the case of $k_BT=0.005E_r$, demonstrating the convergence of QMC calculations at sufficiently low temperatures.

Therefore, we conclude that the temperature $k_BT=0.005E_r$ in the main text is low enough such that it reflects the zero-temperature properties of the superfluid fraction and Leggett's bound.

\end{document}